\documentclass[%
 reprint,
 superscriptaddress,
 onecolumn,
 amsmath,
 amssymb,
 10pt,
 aps,
 pra,
 citeautoscript,
 notitlepage,
]{revtex4-1}

\usepackage{graphicx}
\usepackage{dcolumn}
\usepackage{array}
\usepackage{bm}
\usepackage[colorlinks=true,allcolors=blue,breaklinks=true]{hyperref}
\usepackage{breakcites}
\usepackage[utf8]{inputenc}
\usepackage{xcolor}

\begin{document}

\title{Finding Multiple Reaction Pathways of Ligand Unbinding}

\author{Jakub Rydzewski}
\email[To whom the correspondence should be addressed: ]{jr@fizyka.umk.pl}
\affiliation{Institute of Physics, Faculty of Physics, Astronomy and 
Informatics, Nicolaus Copernicus University, Grudziadzka 5, 87--100 Torun, 
Poland}

\author{Omar Valsson}
\affiliation{Max Planck Institute for Polymer Research, Ackermannweg 10, 
D-55128 Mainz, Germany}

\begin{abstract}
Searching for reaction pathways describing rare events in large systems 
presents a long-standing challenge in chemistry and physics. Incorrectly computed 
reaction pathways result in the degeneracy of microscopic configurations 
and inability to sample hidden energy barriers. To this aim, we present 
a general enhanced sampling method to find multiple diverse reaction pathways 
of ligand unbinding through non-convex optimization of a loss function 
describing ligand-protein interactions. The method successfully overcomes 
large energy barriers using an adaptive bias potential, and constructs 
possible reaction pathways along transient tunnels without the initial guesses 
of intermediate or final states, requiring crystallographic information only. 
We examine the method on the T4 lysozyme L99A mutant which is often used as 
a model system to study ligand binding to proteins, provide a previously 
unknown reaction pathway, and show that using the bias potential and the tunnel
widths it is possible to capture heterogeneity of the unbinding mechanisms 
between the found transient protein tunnels.
\end{abstract}

\maketitle


Molecular dynamics (MD) simulations provide sufficient temporal and spatial 
resolution to study physical processes. Unfortunately, MD fails to reach high 
energy barriers ($\gg k_{\mathrm{B}}T$) that dictate mechanisms and 
kinetics of rare events. 
Transport in heterogeneous media, such as ligand unbinding, cannot be simulated 
directly, and even biased MD methods often fail to find possible reaction 
pathways along complex transient tunnels of proteins 
that form spontaneously during
dynamics~\cite{rydzewski2017ligand,rydzewski2017rare,bruce2018new}. 
Although many general purpose methods have been developed to sample rare 
events~\cite{grubmuller1995predicting,voter1997hyperdynamics,laio2002escaping}, 
finding multiple reaction pathways of ligand unbinding is especially difficult. 
Also, experimental methods used currently to quantify ligand binding, e.g., 
time-resolved crystallography and xenon binding focus primarily on gaseous 
species, providing indirect evidence for the migration of larger ligands, which 
makes most details of reaction pathways unresolved. 

The main computational limitations that render the reconstruction of reaction 
pathways for ligand unbinding difficult stem from accounting for internal 
topological features of proteins (e.g., tunnels), which is related to the 
degree of coupling between protein dynamics and ligand conformational states. 
The structural flexibility of protein tunnels allows proteins to facilitate 
binding by adapting to binding partners along possibly multiple pathways to the 
binding site. This intrinsic dynamics poses a severe challenge to 
straightforward biased MD methods that have been used to sample reaction 
pathways in ligand unbinding~\cite{ludemann2000substrates,laio2002escaping,
miao2015gaussian,wang2016mapping}. 
Typically, such methods either approximate reaction pathways by 
linear Cartesian coordinates~\cite{heymann2000dynamic}, or probe protein tunnels 
randomly~\cite{ludemann2000substrates,kokh2018estimation}.

An additional and ubiquitous obstacle in describing ligand unbinding is the 
overestimation of energy barriers, and thus, the underestimation of exponentially 
dependent kinetic rates arising sampling crude reaction pathways. In other words, an 
inadequate initial guess of reaction pathways leads to false thermodynamics and
kinetics. Another problem which is related to the degeneracy of microscopic configurations
originating from inability to capture intrinsic degrees of freedom, which is
likely to shadow hidden energy barriers~\cite{schneider2017stochastic,zhang2018unfolding}.  
As emphasized by Elber and Gibson~\cite{elber2008toward}, sampling 
should not overestimate preference to more direct and geometrically shorter 
reaction pathways. Producing and exploring multiple reaction pathways of a 
complex system remains a huge challenge~\cite{valsson2016enhancing}.

In this Letter, we consider a specific part of ligand binding/unbinding problem
that is very relevant and not yet fully solved~\cite{ribeiro2018kinetics}. 
To our knowledge, this is the first work to show that sampling multiple transient 
ligand tunnels in proteins leads to heterogeneous mechanisms of unbinding
between the sampled reaction pathways. We present a general enhanced sampling 
MD method to find multiple diverse reaction pathways of ligand unbinding 
along transient protein tunnels. The method does not require many parameters and 
and does not require initial guesses of intermediate states~\cite{heymann2008pathways,
templeton2017rock}, which is a major challenge 
for existing methods. Its only prerequisite is the knowledge of the initial 
bound state, without requiring the initial reactive trajectory. The method also 
takes into account protein dynamics, which is important to observe transient 
tunnels.

\begin{figure}
\includegraphics[width=0.7\columnwidth]{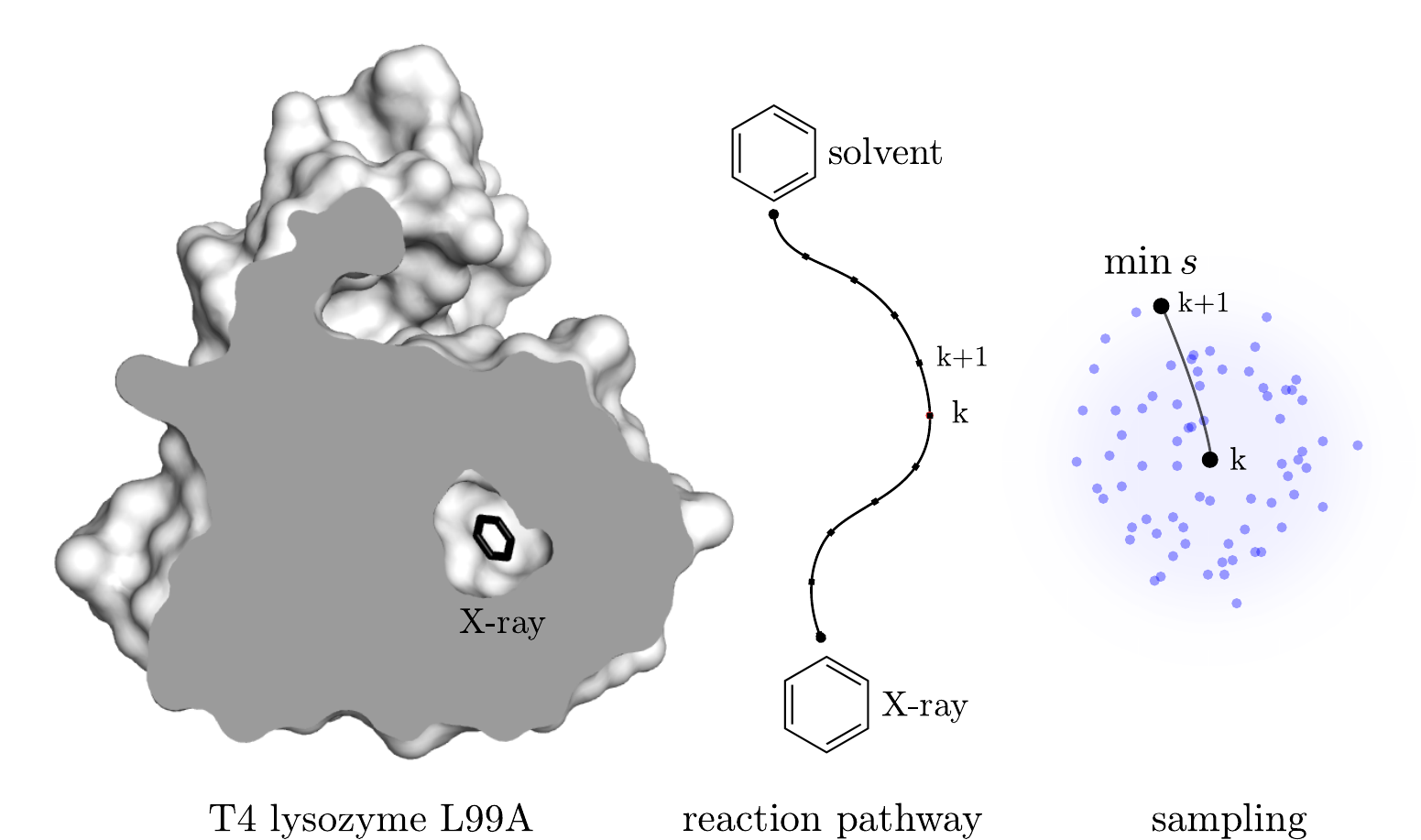}
\caption{Sampling of ligand unbinding pathways using the presented biased MD 
method. As an example, the unbinding of benzene from T4 lysozyme L99A along a 
reaction pathway is shown. The unbinding is initiated from the bound state 
(X-ray binding site) of the T4L-benzene complex, and ends once the ligand 
reaches solvent. (a) The cross-section through X-ray structure of T4L shows no 
apparent tunnels for benzene to leave the protein, which means that the protein 
must undergo structural changes to open possible exits. (b) A 
reaction pathway characterizing atomistically the unbinding along the transient 
exit tunnel is identified locally during the MD simulations. (c) Namely, to determine
the $(k+1)$th intermediate, the conformations of benzene are sampled in the 
neighborhood of the $k$th intermediate (constrained by the sampling radius).  
Then, from the sampled ligand conformations the optimal direction of biasing is 
calculated by selecting the ligand conformation which has the lowest loss 
function score.}
\label{fig:0}
\end{figure}

To estimate ligand-protein interaction we introduce the concept of a loss
function. The method minimizes a loss function $s({\bf x}, {\bf y})$ 
during MD simulations of a $3X$ set of ligand coordinates ${\bf x}\equiv(x_1, \dots, 
x_{3X})$ and a $3Y$ set of protein coordinates ${\bf y}\equiv(y_1, \dots, 
y_{3Y})$, where $X$ and $Y$ are the numbers of ligand and protein atoms,
respectively. To this aim, we propose the loss function must fulfill three 
important criteria, e.g., (i) describe physical interactions in a 
ligand-protein system, (ii) tend to infinity as the ligand moves too close to 
the protein, and (iii) decrease as the ligand unbinds from the protein; (ii) 
prohibits the method from sampling ligand configurations that clash with a 
protein, and (iii) provides a coarse estimate of how ligand conformations are 
buried within a protein tunnel. 

For a schematic depiction of the method, see Fig.~\ref{fig:0}. The method 
follows a procedure: (i) it finds a minimum of the chosen loss function 
in the neighborhood of the current position of the ligand, and, 
(ii) the position of the ligand is biased in the direction of the localized 
minimum of the loss function. The minimization is repeated during 
the MD simulation every $\Delta t$ MD steps, and the biasing is performed 
until a new solution in the neighborhood of the current position is 
calculated. In what follows, we explain in detail the above general outline. 

We start off by describing the loss function and the 
minimization procedure which provides a possible ligand configuration sampled 
in the proximity of the current ligand conformation from the MD simulation, 
which corresponds to 
the lowest loss function score, and by explaining how the neighborhood is 
defined for such an optimization problem. Next, we move on to the adaptive 
biasing procedure which enforces the ligand conformation to dissociate toward 
the conformation selected by the minimization procedure. The method is then
summarized, and used to model the benzene unbinding reaction pathways from the
T4 lysozyme L99A mutant that is often used as a model system to study ligand
unbinding processes.


\textit{Loss Function}.---Because empty space in proteins and its intrinsic 
fluctuations constitute a key 
feature of tunnels~\cite{rydzewski2017ligand,bruce2018new}, we use a coarse 
physical model for ligand-protein interactions, which accounts for steric effects 
only. We motivated our decision by the simplicity of this approach. For the $i$th pair of 
ligand-protein atoms, we define a partial loss function as
$\frac{\text{e}^{-r_i}}{r_i}$, 
where the rescaled distance between the atoms is given by $r_i=\lambda\|x_k-y_l\|$. 
The $\lambda$ constant sets length scale in the loss function, and is equal to 
1 when using ångströms, i.e., $\lambda=1~\text{\AA}^{-1}$. Hence, we used the loss 
function of the following form:
\begin{equation}
\label{eq:1}
  s=\sum_{i=1}^{P_l}\frac{\text{e}^{-r_i}}{r_i},
\end{equation}
where $P_l$ is the number of ligand-protein atom pairs in the local neighborhood
of the ligand (see Supporting Information for details). The sum over all pairs
meets the criteria of the loss function for ligand unbinding presented here. 
The aim of the proposed method is to efficiently sample the configurational 
space of the ligand-protein complex, and optimize Eq.~\ref{eq:1} during MD 
simulations, so that the reconstructed reaction pathways of ligand unbinding 
minimize the loss function along multiple tunnels. We also allow the ligand
to be flexible during the unbinding simulations. In this method, many MD 
simulations are required to sample multiple reaction pathways.


\textit{Minimization}.---Such an optimization problem can be solved by
any method suitable for non-convex loss functions~\cite{hansmann2002global,
rydzewski2015memetic,rydzewski2018conformational}. Here, for the sake of
simplicity the minimization of the loss function is performed using simulated 
annealing~\cite{kirkpatrick1983optimization}. To this end, the method checks if 
a randomly chosen neighboring position of the ligand ${\bf x}'$ is preferred in 
terms of the loss function. The neighbor is selected as a next solution 
according to the Metropolis-Hastings algorithm~\cite{metropolis1953equation} 
with the Boltzmann factors (we omit protein coordinates ${\bf y}$
only in notation): 
\begin{equation}
  p=\begin{cases}
    \text{e}^{-\beta_j\big(s({\bf x}')-s({\bf x})\big)} & \text{if $s({\bf x}')>s({\bf x})$},\\
    1 & \text{otherwise},
  \end{cases} 
\end{equation}
where $\beta_j=1/T_j$ is a parameter introduced to decrease the probability of 
acceptance of a worse solution as the minimization scheme proceeds. $T_j$ 
is reduced according to the recursive formula, $T_j=kT_{j-1}$, where $j$ is the iteration number 
during the optimization phase, to promote convergence to an 
optimum~\cite{kirkpatrick1984optimization}. The minimization procedure is 
reiterated to find an optimal solution. For details concerning the parameters
for simulated annealing, see Supporting Information.

Next, we describe how the neighborhood is defined in our method.
The minimization procedure needs constraints to optimize the 
loss function locally (in the current neighborhood). In our method, intermediate 
ligand unbinding states are searched for sequentially to get an optimal
transition between the X-ray structure and the unbound state. 
A global minimization of the loss function without a specific definition of 
the neighborhood would identify only the final state of ligand unbinding. 
A naive approach~\cite{rydzewski2015memetic,rydzewski2016machine,
rydzewski2017thermodynamics} is to sample ligand conformations constrained 
to a sphere with a constant radius, and positioned at the center of mass of the 
ligand, but this requires an estimate of the radius, which is clearly 
system-dependent and should changed as protein dynamics is simulated. 
To alleviate this issue, we take the sampling radius equal to the minimal 
distance between the ligand-protein atom pairs, e.g., $r_s=\min_{i}r_i$. By 
doing so, the method dynamically adjusts the conformational space available 
for the sampling. We underline that the protein neighborhood of the ligand 
changes as the ligand dissociates during the simulation, and so does the number of
ligand-protein atom pairs $P_l$, which makes the identification of the next
minimum possible.


\textit{Adaptive Biasing}.---Once the optimal ligand-protein conformation 
${\bf x}'=\min_{\bf x}s({\bf x})$ is calculated in the minimization scheme, the 
conformation of the ligand is biased in 
the direction of ${\bf x}'$ along transient protein tunnels. This stage is 
performed by biasing the conformation of the ligand using an adaptive harmonic 
potential:
\begin{equation}
\label{eq:2}
  V({\bf x})=\alpha\left(v\Delta 
  t-({\bf x}-{\bf x}'_{i-1})\cdot\frac{{\bf x}'_i-{\bf x}'_{i-1}}{\|{\bf
  x}'_i-{\bf x}'_{i-1}\|}\right)^2,
\end{equation}
where ${\bf x}'_i$ is the $i$th optimal solution, $v$ is the biasing rate, $\Delta 
t$ is the MD time between subsequent loss function minimizations, and $\alpha$ is the 
force constant. The bias potential (Eq.~\ref{eq:2}) is a generalization of 
the harmonic biasing potential introduced by Heymann and 
Grubm\"{u}ller~\cite{heymann2000dynamic} to curvilinear reaction pathways. 
The biasing potential from Ref.~\onlinecite{heymann2000dynamic} uses a constant
direction of biasing, but in Eq.~\ref{eq:2} this direction is approximated as
the normalized difference between the subsequent minima of the loss function.
In contrast to this method, several recently introduced approaches used a 
constant bias to sample complex reaction pathways~\cite{rydzewski2015memetic,
rydzewski2016machine,kokh2018estimation}. The bias potential shown by 
Eq.~\ref{eq:2} is adaptive, and dependent on the optimal reaction pathways 
calculated by minimizing Eq.~\ref{eq:1}. 

The harmonic bias potential used 
in this study is selected to be simple as the potential energy in MD simulations
already includes bonded terms for interaction of atoms that are linked by
covalent bonds, and nonbonded terms that describe long-range electrostatics and
van der Waals forces. Clearly, the bias potential should not be expected to be quantitative 
as a method to calculate energy barriers along the reaction pathways, 
but it was employed to enforce the process of ligand unbinding with a constant
velocity as in Ref.~\onlinecite{heymann2000dynamic}. The bias potential, however,
may serve as a means to shorten the time-scale of ligand unbinding, 
and as a qualitative measure to estimate relative differences of
bias between the reaction pathways.

Our enhanced sampling method is outlined as follows:
\begin{enumerate}
  \item Initialize the MD simulation,
  \item Sample ligand conformations within the protein tunnel using constraints
    defined as the minimal distance between the ligand-protein atom pairs,
  \item Minimize the loss function using a non-convex optimization algorithm and
    set the biasing direction toward the found minimum,
  \item Bias the ligand conformation using Eq.~\ref{eq:2} during $\Delta t$
    steps of the MD simulation,
  \item Repeat the steps 2--4 during the MD simulation until the loss function
    reaches zero,
  \item Stop the MD simulation,
\end{enumerate}
which concludes the introduction of the method components, e.g., loss function,
minimization, and adaptive biasing, needed to sample ligand unbinding reaction
pathways.


\textit{Unbinding Benzene from T4 Lysozyme L99A}.---We illustrate the method on 
T4 lysozyme L99A (T4L) with bound benzene, which is considered as a model 
system to study ligand unbinding from proteins. In this example, 300 10-ns 
trajectories were run to reconstruct the reaction pathways of benzene unbinding 
from the protein. We used the biasing rate $v=0.02$ \AA/ps with the force 
constant in the stiff-spring regime~\cite{heymann2000dynamic}, $\alpha=3.6$ 
kcal/(mol\,\AA). The optimal position of the ligand was recalculated by 
minimizing the loss function every $\Delta t=200$ ps. We found that for lower 
biasing rates the method is unable to find the reaction pathways in the desired 
span of 10 ns for a single simulation. This is, 
however, only a technical nuisance that can be overcome 
by sampling longer MD trajectories. The method is implemented in the 
official Plumed-2.5 repository~\cite{tribello2014plumed} which is  
available on Github~\cite{maze} and described in Ref.~\onlinecite{rydzewski2019maze}.
For details concerning the model of T4L with bound benzene and the MD 
simulations, see Supporting Information.

\begin{figure}
\includegraphics[width=0.7\columnwidth]{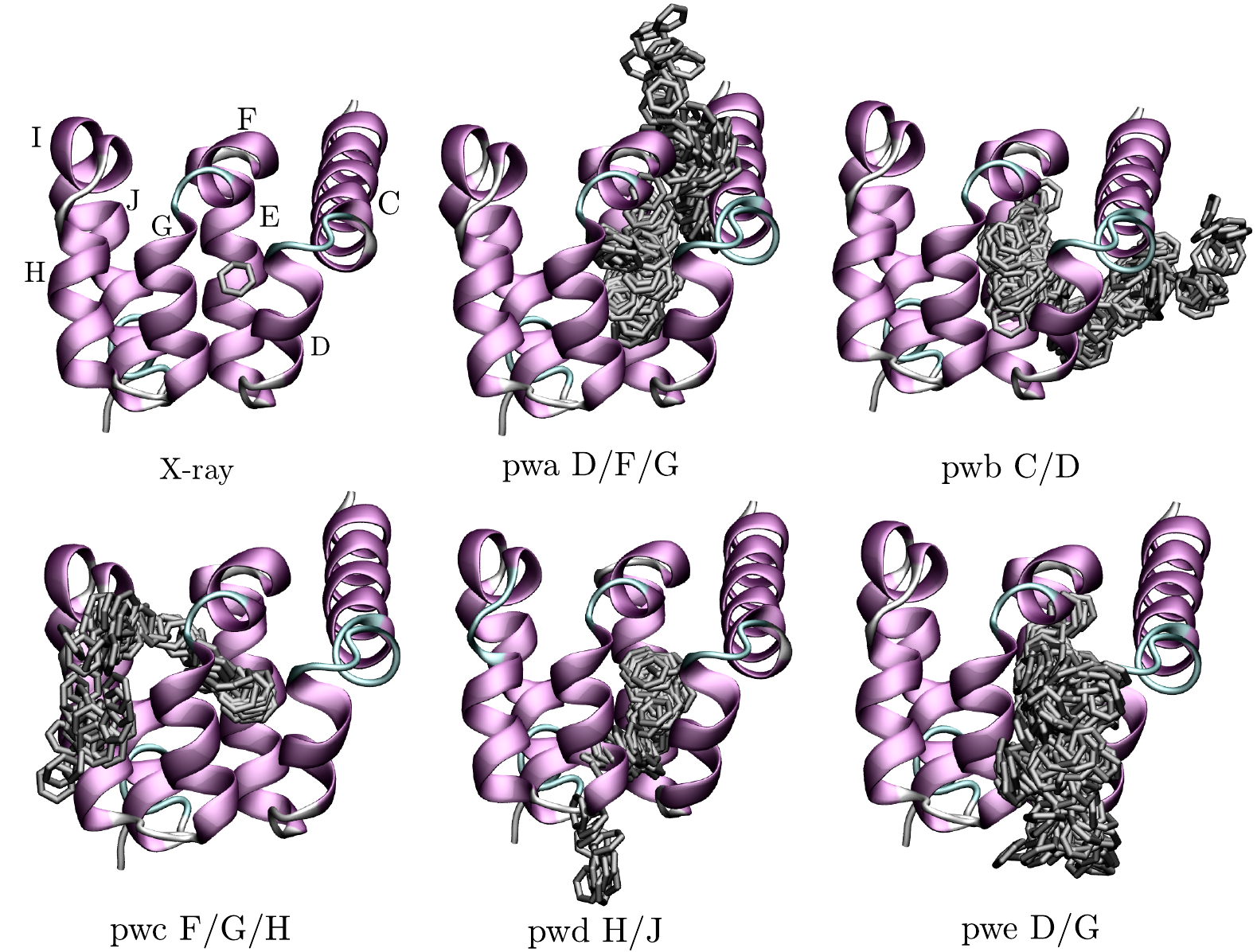}
\caption{Reaction pathways of benzene unbinding from T4L. Only the T4L 
C-terminal domain is depicted, but the complete protein was used in all 
simulations. The crystallographic bound conformation of benzene is shown. 
Benzene conformations sampled during the MD simulations are biased by the 
adaptive bias potential to find multiple exits of T4L via the reaction 
pathways. The reaction pathways are named pwa-e, which corresponds to the T4L 
tunnels indicated by helices, i.e., D/F/G tells that the unbinding pathway is 
located near the D, F, and G helices.
\label{fig:1}}
\end{figure}

We directly compared our results with reaction pathways found in previous 
studies. The method identified five reaction pathways for benzene exit from the 
binding cavity buried in T4L. These reaction pathways correspond to five 
tunnels of T4L, named pwa--D/F/G (tunnel through helices D, F, and G), 
pwb--C/D, pwc--F/G/H, pwd--H/J, and pwe--D/G (Fig.~\ref{fig:1}). The 
reconstructed reaction pathways pwa--d are mostly in agreement with a recent 
study by Nunes-Alves et al.~\cite{nunes2018escape} in which the 
reaction pathways of benzene unbinding were sampled 
using temperature-accelerated MD simulations~\cite{nunes2018escape}. Other 
studies also reported pwa~\cite{miao2015gaussian} and pwc~\cite{wang2016mapping}. 
To our knowledge, the benzene unbinding via pwe is first identified in this
study.

Apart from the work of Nunes-Alves et al.~\cite{nunes2018escape}, other studies 
found only one reaction pathway, probably because of the employed biased MD 
methods. Biased MD methods employed by Wang et al.~\cite{wang2016mapping} and 
Miao et al.~\cite{miao2015gaussian} may limit the search in configuration space 
to a most optimal solution. Wang et al. used 
metadynamics~\cite{barducci2008well} to bias a reaction pathway identified 
initially by self-penalty walk~\cite{nowak1991reaction}, which agrees with the 
observation that such methods strongly rely on the initial guess of a 
pathway~\cite{passerone2001action,lee2017finding}. Thus, it may not be possible 
to identify all possible reaction pathways that exist in the form 
of transient sparse tunnels in the studied ligand-protein complex. 
Interestingly, the reaction pathways identified here agree mostly with exit 
tunnels retrieved by Nunes-Alves et al.~\cite{nunes2018escape}, where MD 
simulations with elevated temperature were used to overcome large energy 
barriers along reaction pathways and increase the probability of the rare 
event. In both temperature-accelerated MD~\cite{abrams2010large}
and the method presented in this article there is no need for initial guess 
of trajectories, which clearly improves sampling of diverse pathways.

\begin{table}[h]
\caption{Reaction pathways of benzene unbinding from T4L. Quantities describing 
the reaction pathways were calculated from an ensemble of trajectories for the 
identified exits. These quantities include the number of trajectories (out of 
300) that proceed through each pathway, the mean of the distribution of 
unbinding times that it takes for benzene to unbind in the biased simulations, 
and its standard deviation, and the average radius of each identified tunnel 
$r_s$ and its standard deviation. Errors were estimated by a bootstrapping 
procedure (see Supporting Information).}
\label{tab:1}
\begin{tabular}{llcccc}
\hline
	pathway & tunnel & no. trajectories & unbinding time [ns] & $r_s$ [\AA]\\
\hline
	pwa & D/F/G & 65 & $3.37\pm0.01$ & $2.37\pm0.01$\\
	pwb & C/D   & 82 & $2.61\pm0.07$ & $2.31\pm0.01$\\
	pwc & F/G/H & 34 & $2.68\pm0.09$ & $2.41\pm0.02$\\
	pwd & H/J   & 27 & $2.29\pm0.08$ & $2.36\pm0.03$\\
	pwe & D/G   & 92 & $2.45\pm0.11$ & $2.34\pm0.01$\\
\hline
\end{tabular}
\end{table}

\begin{figure}
\includegraphics[width=\columnwidth]{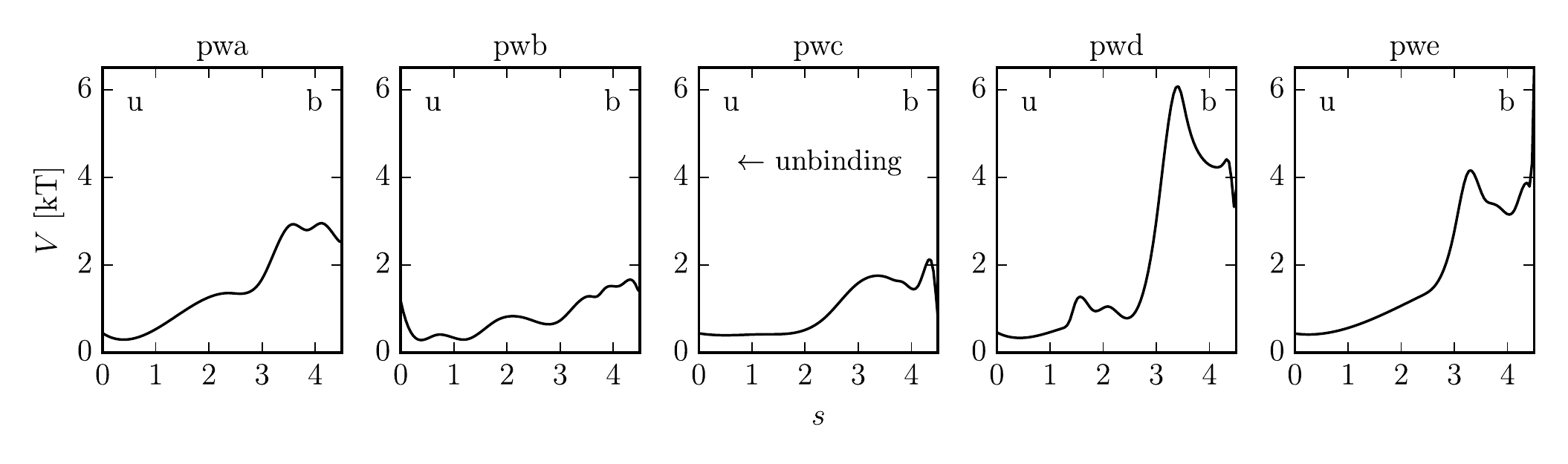}
\caption{Averaged bias potentials $V(s)$ from all the simulations that took a specific pathway 
projected along the reconstructed reaction pathways. Here, we used the loss 
function $s$ to project the bias potential to depict in what stage of 
unbinding the bias is higher. The high value of the loss function indicates 
that the ligand is bound (b) to the T4L matrix (in the X-ray structure the 
loss function reaches about 4.5), whereas the low value an unbound (u) state 
(at end of MD simulations the loss function decreases to 0). As can be seen, 
the characterization of the reaction pathways is heterogeneous between the
different classes, showing different mechanisms of the benzene unbinding, and 
indicates different bias potential barriers for the reaction pathways close 
to one another, for instance, pwa and pwe near the D helix. 
\label{fig:3}}
\end{figure}

The detailed characteristics of the reaction pathways for benzene unbinding 
from T4L are shown in Tab. I. We found an additional reaction pathway that, to 
the best of our knowledge, was not identified previously. This pathway 
corresponds to the benzene unbinding along the T4L tunnel between helices D and G. 
The method is able to provide the atomistic characterization of unbinding pathways.
If, however, one is interested in knowing estimates of the energy barriers 
along unbinding pathways, a postprocessing procedure is needed to analyze the data.
For instance, one way to further understand the results is to look at the 
averaged bias potential $V(s)$. We followed this approach, and average the 
bias potential along each pathway projecting it on the loss function (Fig.~\ref{fig:3}) 
using the following relation:
\begin{equation}
  V(s)=\bigg\langle\delta(s-s({\bf x}))V({\bf x})\bigg\rangle_p,
\end{equation}
where the average $\left<\cdot\right>_p$ is taken over all trajectories
classified as a particular reaction pathway $p$, and $s$ is the loss
function defined in Eq.~\ref{eq:1}.
This way we were able to identify energy bottlenecks in tunnels indicated by high 
values of the averaged bias potential $V(s)$ or the sparsity of conformational 
space available for sampling. Treating the loss function as a collective variable, 
although may be not intuitive, provides a simple formula to check which 
transient tunnels are biased the most. Despite roughly the same level of the 
bias along each reaction pathway (Fig.~\ref{fig:3}), it is clear that the 
pathways employ different mechanisms of unbinding, without the need to 
reconstruct free energies. This is underlined by the bias barriers along pwd and pwe, 
and a rather smooth decrease of the bias along pwa, pwb, and pwc. 
Moreover, it is perhaps possible to explain 
the bias barriers by inspecting the average radius of each tunnel which is used
as the sampling radius $r_s$. For instance, pwd and pwe have $r_s$ at about 2.36
~\AA~and 2.34~\AA, respectively, and the highest barriers among pathways. This 
is an indication that the reaction pathways are heterogeneous with respect to 
each other, and their specific atomistic mechanism of unbinding would not be 
obvious by calculating averages of the full ensemble of the unbinding 
trajectories, without decomposition into classes first. 

As recently underlined in Ref.~\onlinecite{feher2019mechanisms},
multiple pathways for benzene escape out
of the T4L crystallographic binding site exist if one of the end states consists
of multiple substrates~\cite{bhatt2011beyond}. 
Interestingly, the reaction pathway via F/G/H tunnel
identified by Fehler et al.~\cite{feher2019mechanisms} is argued 
to be the most probable in their
study. Our results show that this result may be due to the highest tunnel width
in comparison to the other pathways (Tab.~\ref{tab:1}). As it is shown in
Tab.~\ref{tab:1}, the sampled trajectories that yield the same reaction pathways
(same tunnels) are similar to each other as it is underlined by the small
standard deviations of the unbinding time and the sampling radius estimated
using bootstrapping (see Supporting Information).

It should be noted that the method lends itself to use as an optimal initial 
guess of reaction pathways in other biasing MD methods to estimate 
thermodynamic and kinetic quantities, i.e., metadynamics~\cite{laio2002escaping,
tiwary2013metadynamics} or variationally enhanced sampling~
\cite{valsson2014variational}. We point out 
that computing reaction pathways for the T4L-benzene complex is not needed 
when calculating the mean-first-passage times of binding and unbinding as shown 
by Wang~\cite{wang2018frequency}, however, it is important in estimating how the 
mechanisms of binding varies between the calculated reaction pathways, 
including free energies and conformational changes. Recently, it was shown that 
some protein-ligand systems can exhibit pathway 
hopping~\cite{rydzewski2018kinetics,lotz2018unbiased}, and the method presented 
here can be used to quantify this process.   

We note that the reaction pathways of ligand unbinding sampled using the method 
presented here diverge to diverse suboptimal basins. This is the feature that 
enables sampling multiple heterogeneous reaction pathways and allows to 
overcome the problem of the intrinsic dynamics of protein tunnels. This is due 
to the used sampling which is constrained by the protein structure to provide a 
local minimum. Also, the probability of selecting a new solution given by the 
Metropolis-Hastings algorithm is important for the 
heterogeneity of the reaction pathways. The method searches for an optimal 
ligand conformation locally to extend the current reaction pathway step by 
step. This way, the method is able to sample multiple possible unbinding 
pathways, which for a rare event as with ligand unbinding is necessary to 
explore configurational space of tunnels exhaustively.


In conclusion, we have presented a general method for finding reaction pathways 
of ligand unbinding, starting only from available crystallographic information. 
The method does not need any prerequisite guesses of intermediate  
states. The introduced approach uses an adaptive bias to drive the ligand to 
unbind from the fluctuating protein, in the direction effectively calculated by 
minimizing a simple loss function. The methods adapts to transient tunnels of 
proteins by estimating the configurational space from which it samples 
plausible ligand conformations (i.e., it can be also used to determine the
tunnel widths). We think the method should be applicable to 
proteins in which prominent structural motions on a larger scale are important 
for ligand unbinding (e.g., trypsin~\cite{tiwary2015kinetics}). 

Various enhanced sampling techniques have been tested for characterization of
rare events and long-timescale dynamics. The method proposed here was suitable
to sample a rare conformational event such as benzene escape that occurs on the
millisecond timescale experimentally. We provided a rigorous method to find 
possible reaction pathways, which can be used as a 
initial reference trajectory to reconstruct thermodynamic and kinetic data. 
Overall, our results from 
studies of ligand unbinding from T4L suggests that the method presented here 
can improve the reconstruction of reaction pathways along 
transient tunnels, and serve as an optimal choice for other biasing methods, 
limiting overestimation of hidden free energy barriers. With some
adaptations the method can be 
also used to study other transport processes, e.g., diffusion through 
a membrane.

{\bf Note.} At submission stage we became aware of Ref.~\onlinecite{capelli2019exhaustive}
in which the benzene unbinding pathways from the T4L protein are also
studied. Capelli et al. found various reaction pathways that they classified
into eight reaction pathways using a different criteria than we use here. In
particular, the benzene unbinding pathways marked by Capelli
et al. as C, F, G, and H~\cite{capelli2019exhaustive} are subclasses of pwc,
while other four are the same as ones we have identified here. 

\section*{Supplementary Material} 
See supplementary material for: model of T4 lysozyme L99A; MD simulations; 
loss function; neighborhoodfor  the  loss  function;  minimization  procedure;  
adaptive  biasing  to  a  loss  function  minimum;  classification  of  the
reaction pathways; biased unbinding times; software.


\begin{acknowledgments}
We thank H. Grubm\"{u}ller, W. Nowak, and M. Parrinello for useful discussions, 
and Tristan Bereau and Claudio Perego for critically reading the manuscript. 
This work was supported by the National Science Centre, Poland (grant 
2016/20/T/ST3/00488, and 2016/23/B/ST4/01770). The MD simulations were computed 
using facilities of 
Interdisciplinary Centre of Modern Technologies, Nicolaus Copernicus 
University, Poland.
\end{acknowledgments}

\bibliography{lb99}

\end{document}